**The Communication of Meaning and the Structuration of Expectations:**

**Giddens' "structuration theory" and Luhmann's "self-organization"**



Loet Leydesdorff

Amsterdam School of Communications Research (ASCoR), University of Amsterdam,

Kloveniersburgwal 48, 1012 CX Amsterdam, The Netherlands;

loet@leydesdorff.net ; http://www.leydesdorff.net

**Abstract**

The communication of meaning as different from (Shannon-type) information is central to Luhmann's social systems theory and Giddens' structuration theory of action. These theories share an emphasis on reflexivity, but focus on meaning along a divide between inter-human communication and intentful action as two different systems of reference. Recombining these two theories into a theory about the structuration of expectations, interactions, organization, and self-organization of intentional communications can be simulated based on algorithms from the computation of anticipatory systems. The self-organizing and organizing layers remain rooted in the double contingency of the human encounter which provides the variation. Organization and self-organization of communication are reflexive upon and therefore reconstructive of each other. Using mutual information in three dimensions, the imprint of meaning processing in the modeling system on the historical organization of uncertainty in the modeled system can be measured. This is shown empirically in the case of intellectual organization as "structurating" structure in the textual domain of scientific articles.

**Introduction**

Shannon (1948, at p. 379) detached himself from the communication of meaning by stating on the first page of his mathematical theory of communication that "these semantic aspects of communication are irrelevant to the engineering problem." His co-author Weaver, however, noted that "this analysis has so penetratingly cleared the air that one is now, perhaps for the first time, ready for a real theory of meaning" (Shannon & Weaver, 1949, at p. 117). The communication of meaning as different from information is central to the sociological enterprise in which one attempts to explain how human action is socially coordinated (Berger & Luckmann, 1966; Schutz [1932] 1967). In the sociology of science, the communication of knowledge in scholarly discourses presumes that information and meaning can further be codified by being communicated (e.g., Bell, 1973, at p. 20; Dasgupta & David, 1994; Gilbert, 1997; Leydesdorff, 2007; Mulkay *et al.*, 1983).

In this study, I focus on the operationalization of the communication of meaning. First, I shall address the conceptual issues and then show that meaning as interhuman coordination mechanism can be considered as anticipatory (Rosen, 1985). In my opinion, Luhmann's sociological theory of communication can be operationalized by using Giddens' structuration theory if the latter is understood as a theory about the structuration of expectations instead of action. The computation of anticipatory systems (Dubois, 1998) allows us to develop simulation models for the communication of meaning at three different levels. These levels correspond to Luhmann's (1986) elaboration of Husserl's



(1929) concept of "intersubjectivity" in terms of (*i*) interactions, (*ii*) organization, and (*iii*) self-organization of communications at the level of society.

One can provide meaning to the events from the perspective of hindsight, that is, against the arrow of time. In other words, a model provides meaning to the modeled system among possible meanings and thus a redundancy can be generated (Krippendorff, 2009a). The model can contain a prediction of a future state of the modeled system. In the third part of this study, I turn to the operationalization of this redundancy using the textual domain of relations among scientific articles as data. Can the communication of meaning be indicated, for example, as intellectual organization that feeds back on the ongoing flow in the communication of information?

**Reflexivity in inter-human communications**

In order to explain the difference between "structure" as a property of social networks at each moment of time and a dynamic conceptualization of structure that facilitates and constraints communicative intent in interhuman interactions, Giddens proposed the concept of "structuration" (1976, at p. 120) and related this concept from its very origin to the double hermeneutics operating in intentional interactions among humans beings. A double hermeneutics is possible in interhuman communication because one can understand someone else as another participant in the communication in addition to observing and interpreting the behavior of the other.



This difference between "action" as an observable practice *versus* "interaction" based on intersubjective understanding can be traced back to Weber's *Economy and Society* ([1922], 1978). However, Schutz ([1932] 1967, at p. 8) noted that Weber had not sufficiently elaborated on the distinction between human action as a unit providing meaning and the interpretation of meaning as a cultural object. Husserl's phenomenological critique of the positive sciences (1929, 1936) and Parsons' concept of "double contingency" (1951, 1968; cf. Mead, 1934) provide relatively independent sources of what can be considered as fundamental background to the sociological enterprise: inter-human interactions can be expected to have both a practical component and communicative meaning (Habermas, 1981).

While all human practice is embedded in a structured historicity, the intentional part is not structured, but structurated: it includes and constitutes different time horizons of meaning by enabling and constraining further actions and expectations. For Giddens (1976), language and semiosis within language provided the model for this theory of "structuration." Structuration was defined by Giddens (1979, at p. 66) as the conditions governing the continuity and transformation of structures, and therefore the reproduction of systems.

While structures can thus be considered as properties of social systems, systems were defined by Giddens (1979, at p. 66) as "reproduced relations between actors or collectivities as regular social practices." Systems, in other words, are instantiated in observable networks of relations (Giddens, 1984). The "duality of structure" is then



proposed as a recursive operation which transforms aggregates of action into systems by invoking structuration as a "virtual" operation. Whereas structures can be analyzed as latent properties of communication systems (at each moment of time), structuration transforms both actions and structures over time by providing them with reconstructed meaning.

Without any references to Giddens, but based on the same sources—that is, Husserl's supra-individual intentionality and Parsons' double contingency—Luhmann (1984) proposed a theory of social systems in which the communication of meaning is considered as the distinguishing characteristic of a social system. The communication of meaning is structured by processes of codification. Codification can be considered as a self-organized result of communications operating reflexively upon one another. Differentiation among the codes in the communication enables the communication systems and the reflexive carriers of communication to process more complexity.

Using another concept of Parsons (1963, 1968), namely, symbolically generalized media of communication, Luhmann further proposed that these symbolic generalizations can be operationalized in terms of functionally different codes of communication. For example, political discourse is coded differently from scientific discourse and both codes are different from the communication codes operating in economic exchange relations or intimate relationships. In modern societies, the codes have been symbolically generalized at the level of society and can reflexively be instantiated by individuals and organizations.



Unfortunately Luhmann (1975a, 1984) used a biological metaphor—like DNA as the code that determines reproduction (Künzler, 1987)—instead of following Parsons' linguistic understanding of coding. Luhmann's codes are binary (on/off) and consequently the functionally different systems of communication are operationally closed and cannot be commensurate with one another. From this perspective of operationally closed systems, translations among codes of communication are doomed to fail like those among paradigms in Kuhn's (1962) *The Structure of Scientific Revolutions*. Giddens (1976, at pp. 142 ff.) critiqued this biological metaphor as follows:

> The process of learning a paradigm or language-game as the expression of a form of life is also a process of learning what that paradigm is not: that is to say, learning to mediate it with other, rejected, alternatives, by contrast to which the claims of the paradigm in question are clarified. (at p. 144).

Furthermore, Giddens' (1984, at p. xxxvii) "repudiated" Luhmann's "newer version of Parsonian functionalism" because abstract principles are assumed that would govern the development of society for "functional" reasons. One thus loses a perspective on the openness of the development of society for human action and intentful interventions. Biological metaphors suggest that these are merely disturbance terms of otherwise autonomous—or in Luhmann's theory *autopoietic*—developments.

The contradiction between these two approaches can also be considered as another round in the discussion between a structuralist or systems approach *versus* a human-centered focus on interaction. In my opinion, this simple scheme is unfortunate because one can



learn from manoeuvring between the Scylla of Giddens' structuration theory of action and the Charybdis of Luhmann's social systems theory of communication. Whereas Luhmann bracketed human action and intention as analytically outside his systems of social coordination, Giddens could no longer specify the duality of structure as a systems operation because he wished to abstain from reifying an "absent set of differences" (Giddens, 1979, at p. 64; cf. Leydesdorff, 1993).

Yet despite the radical differences between (Giddens') action and (Luhmann's) systems theory, reflexivity prevails in the subject matter of both theories. Giddens' structuration theory focuses on reflexivity as constitutive of human action, while Luhmann's theory asks how reflexivity can be codified at a supra-individual level. This distinction in terms of two types of reflexivity finds its origin in Husserl's (1929, 1936) phenomenological critique of the positive sciences as an insufficient base for developing the social sciences. Observable behavior and facts in the social domain cannot be studied only as data input for analysis, but also as events which could have been shaped differently (Bourdieu, 2004).

### *Cogitantes* and *Cogitata*

Not incidentally, Husserl (1929) entitled his ground-breaking essay *Cartesian Meditations:* Descartes' *Cogito*—"I think"—remains our sole source of reflexivity as agency. The *Cogito* itself operates as a circle: awareness guides us in providing meaning to what we perceive. Specifically, awareness of another *Cogito* can provide meaning to



what can be shared among *cogitantes*—that is, reflexive human beings—and how we may differ in terms of our expectations.

Among other things, *cogitantes* first share a reflexive relation to the *res extensa*, that is, the physical reality. However, what things (e.g., the body) *mean* for each of us may differ greatly. Awareness of the possibility of entertaining different expectations constitutes horizons of expectations that may operate upon us as emerging "intersubjectivity." This intersubjective domain—social order—emerges from interactions among expectations. It enables us, for example, to agree (or disagree) upon transgressions of the order of social expectations. Although partly externalized from the perspective of each individual, this order remains *res cogitans*; Husserl labeled it as a *cogitatum*—the results of the reflections. Note that one remains uncertain about this result.[1] A *cogitatum* can be expected to operate differently from a *cogitans*.

In addition to expectations about our physical and biological realities, we as *cogitantes* are reflexively able to entertain models about the order of expectations among us. Models provide meaning to the modeled systems, and models can be refined in discourse. Luhmann proposed to consider such exchanges of meaning as the proper domain of sociology: How are the *cogitata* structured by interhuman interactions? Note that one has no access to this sociological domain other than in terms of expectations: epistemologically the *cogitata* remain hypotheses which are part of a theoretical discourse that can be entertained by *cogitantes*.

---

[1] For Descartes (1637), the *cogitatum* is transcendent and refers to God as the Other of the *Cogito.* Because this Other is *not* a *Cogito,* but a Transcendency, it can no longer be uncertain.



How can the exchange of expectations about expectations inform us about expectation at the intersubjective level and/or the subjective level? Given the double hermeneutics, the two questions of what specific expectations mean for reflexive individuals and/or for their communication cannot radically be uncoupled because the one hermeneutics is reflexive on the other. However, the systems of reference are different and the relations among them are not necessarily symmetrical. Luhmann suggested that the analytical distinction in terms of two different systems of reference (*cogitantes* and *cogitata*) can clarify the sociological domain as distinct from psychological expectations. Can the order of supra-individual expectations be discussed and explained in a specifically sociological discourse?

Parsons' (1951, 1968) concept of "double contingency" in interhuman communication is used by Luhmann (1995) as the stepping stone for developing a sociological theory about this *cogitatum*: in addition to our awareness of each other, each can expect the other to be reflexive and to entertain expectations just as we do. These expectations are contingent upon one another, but in a domain different from the physical one. In other words, the relations between potentially different *cogitata* contain a second contingency: these interactions among expectations are embedded in a social order and social institutions. Beyond a double contingency in interactions, triple and higher-order contingencies can also be expected (Strydom, 1999).



For example, one can entertain different expectations about others' expectations in private or public configurations. The expectations of third parties matter. Because there is no given order in how the interactions interact, the dynamics of this complex system can be expected to "self-organize" unless an order is imposed. In the latter case, Luhmann (e.g., 1975b, 2000) proposed to consider the communication as organized. Thus, three levels were distinguished: interaction, organization, and self-organization of communications as communication-theoretical elaborations of Husserl's yet insufficiently specified concept of "intersubjectivity" (Luhmann, 1986).

Intentionality is grounded in experience and reflexivity at the individual level. Luhmann (1986) used Maturana and Varela's (1980) concept of "structural coupling" for the necessary relationship between the two layers of "intersubjective" communication and agent-based consciousness: the neural network operates in terms of a distribution of neurons firing, but with a self-organizing (*autopoietic*) dynamic different from those based on recursive interactions among interactions (Maturana, 1978). Unlike the neural network, however, social order is not *res extensa*, but an order of expectations. This next-order level of expectations is not another agency, but remains part of our reflexive experience. Social order not only cannot operate without human agents, but the content of communication additionally matters for subsequent reflections.

In this second contingency—the layer of expectations as different from the contingency among observables—the meaning of communication at the individual level and at the social level are operationally coupled in terms of possible reflections in addition to the



structural coupling between *cogitantes* and *cogitata* as two meaning-processing systems. This operational coupling in the processing of meaning is acknowledged by Luhmann ([1988] 2002) as *interpenetration* (cf. Parsons, 1968), but was not elaborated by him otherwise than by pointing to the evolutionary step of human language as a medium of communication (enabling us to communicate both meaning and uncertainty; cf. Leydesdorff, 2000). However, this additional coupling between the reflexive operations of the two systems brings Giddens' structuration theory of action back on stage because the "self-organizing" system can no longer operate without reflexive agency. In my opinion, this additional coupling in the second contingency makes Luhmann's communication systems "quasi-autopoietic:" *cogitantes* are not only the carriers of *cogitata*, but reflexively they also have access to their substantive content (Collier, 2008).

In summary, Giddens (1984) and Habermas (1987) were right that Luhmann's theory had meta-biological overtones (Leydesdorff, 2006 and forthcoming). Structural coupling can be considered as a biological mechanism: a network system is "plastic" with reference to the distribution of agents at the nodes firing. However, each agent is in this case counted only as on/off, not in terms of *what the communication means*. The biological agents at the nodes have no choice other than on/off in reaction to an update at the network level. The reflexive layer among *cogitantes* and *cogitata*, however, evaluates the communications not only in terms of relative frequencies, but also substantively. *Cogitantes* can do this consciously, while the *cogitata* can be structured by the codes of communication and thus configurational meaning is provided to a communicative event.



As long as the codes of communication are organized hierarchically with, for example, religious and political communication at the top as in a high culture, the system of expectations can be integrated symbolically into a single horizon of meanings. Functional differentiation among the codes breaks this cosmological order. A modern and pluriform society can emerge when different codes of communication are free to operate upon one another at the above-individual level, for example, because of civil liberties warranted by a modern constitution.

Academic freedom, freedom of expression, and the pursuit of happiness enable us to move freely between social coordination mechanisms without the need for *ex ante* synchronization. Yet, one is aware that a religious truth is communicated differently from a scientific one. This self-organizing dynamic is counterbalanced by the need for historical integration into what Giddens would call a system, but what from the Luhmannian perspective can count as only an instantiation or a specific organization of the system. The reproduction of historical instantiations provides a retention mechanism from this evolutionary perspective. However, the codes of communication operate as selective control mechanisms at a level which remains largely beyond the control of individuals. The codes can be considered as the results of interactions among interactions in previous communications of meaning; they structure the communicative horizons for the human agents who are reflexively implied in their historical reconstruction.

In other words, the codes of the communication provide the structural properties of the systems of communication, but they operate differently from historical structures in



networks of communication at each moment of time because the codes reduce uncertainty over time. They structurate present developments from a next-order level that includes the dynamic perspective. Structures are historical and can develop along trajectories. Structurations, however, operate as regimes that can restructure what trajectories mean, that is, from the perspective of hindsight. This latter dynamics is self-organized or, in other words, an unintended outcome. However, this social order of expectations feeds *back* at a next-order—that is, relatively global—level on local instantiations. If this feedback mechanism operates, it can be expected to leave an imprint on historical developments.

For example, the relations among texts in a set of scholarly documents may intellectually be organized. The intellectual organization manifests itself in the textual organization. New contributions to the discourse (submitted manuscripts) may reconstruct the intellectual organization as an order of expectations, and if accepted they can be retained in the textual layer. Three contexts are operating: the initial interaction of the new knowledge claim with the existing order of expectations, the validation of this knowledge claim in the context of justification (at the intellectual level), and the dynamics of how this is organized in the historical realm of scholarly publications and their networks of relations (Fujigaki, 1998; Lucio-Arias & Leydesdorff, 2009). I will return to the example more extensively below.

The model is based on (second-order) cybernetics: the construction is bottom-up, but control emerges top-down. Each reflexive reconstruction at the individual level



necessarily takes part in the reconstruction of social order. Reconstruction can also take place at intermediate levels when communications are organized historically, for example, in institutions. The organization of communication can also be carried by institutional agency (e.g., a research program). However, organization in a given historical instance is one degree less complex than the self-organizing dynamics of communications over time. All forward and feedback arrows are possible, and in this sense the system can be considered as infra-reflexive (Latour, 1988). The research problem, however, is to specify the next-order control mechanisms as theoretically informed hypotheses.

By introducing three layers into the *cogitata*—interaction, organization, and self-organization of communications—Giddens' concept of the duality of structure can be operationalized further. At the bottom, new variants can be constructed or existing variants can be reconstructed. For example, in the case of scientific discourse, a knowledge claim can be made by a scholar submitting a manuscript. The manuscript is evaluated in a process invoking standards in codes of communication above the individual level and compared to the current organization of the "state of the art." If accepted, the knowledge claim is organized into the body of knowledge and the manuscript can become part of the archive by being printed in a journal. A socio-cognitive and textual order is thus reconstructed. The textual order retains the achievements which are incorporated into a cognitive order of expectations that co-constructs expectations in a next round by potentially changing the standards for the evaluation of future submissions.



**Towards measurement and simulation**

I have taken the liberty of modifying Luhmann's model so that it can be made to fit with Giddens' model in order to proceed towards an operationalization. To Giddens' model I added that structured systems of expectations can operate at the supra-individual level as horizons of meaning. Codes of communication structurate the reproduced relations among expectations over time in addition to the structures that the communication networks contain at any given moment.

Furthermore, I added to Luhmann's model that individual expectations and the social order of expectations are not only coupled structurally (as a formal mechanism), but also interpenetrate each other, and that this interpenetration makes the *autopoiesis* (or self-organization) dependent on the reflexivity and communicative competencies at the actor level (Habermas, 1981). The *cogitantes* are not only a (formal) precondition for the *cogitata*, but also shape them operationally because of the prevailing reflexivity. The codes are not given (as in DNA), but remain reflexively under constant reconstruction.

In order to proceed to the operationalization, let me use the specific code of communication in scientific discourse as an example. This code allows scholars to develop discursive knowledge using selections at the network level and as a result of scholarly communications among themselves. Discursive knowledge can be reflected by each participant with communicative competency in the specific domain. Individual reflections can generate new knowledge claims in mutual contingencies with colleagues and interactions among them. These knowledge claims provide interactive variation; the



structures of the discourse in the scientific community organize these claims as contributions; and the structuration by the code can be expected to restructure the discourse. Restructuring may be piecemeal, but sometimes reaches to the paradigm level.

Meaning processing cannot be observed directly because meaning originates and remains *res cogitans* in the second contingency. However, the imprint of meaning processing on information processing can be measured, because providing meaning to the observable variation can be expected to operate as a selection mechanism to *reduce* the prevailing uncertainty. This is formalized in the social sciences, for example, as factor analysis: one can reduce the uncertainty in data by using factor analysis. The factor model reinterprets the data and provides them with new meaning (e.g., factor scores and factor loadings) on the basis of a model.

Providing meaning is a recursive operation, that is, an operation which can be applied to its own outcome: some first-order meanings can later be selected as meanings which make a difference with reference to a latent code in the communication. Analogously, some factor models may be considered as more meaningful than others in the light of theoretical considerations. While the factor model analyzes the complexity in the data in terms of different dimensions at a specific moment, the theoretical update along the time dimension can operate selectively upon the different meanings attributed to the different dimensions distinguished by the factor analysis.



Thus, three levels of meaning processing are involved in the dynamics of meaning (Figure 1). First, observable human actions can be considered as meaningful interactions in inter-human communication. Unlike agents, communications as events in the second contingency cannot be observed directly but must be inferred from observable behavior (Luhmann, 1995, p. 164). In other words, meaning is provided by a (mostly implicit) model. This model *organizes* the communications by relating different information in terms of their meaning. When these meanings can also be exchanged, a next-order model of the components in the organization of meaning can be hypothesized. This next-order model structurates the structures, and thus can add to the reduction of uncertainty in the modeled system.

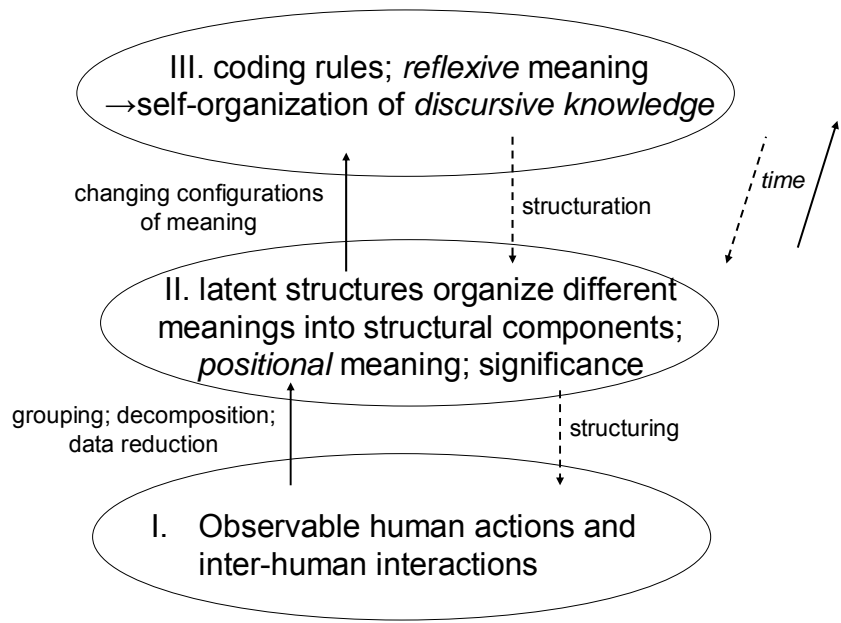

**Figure 1**: A layered process of codification of information by the processing of meaning, and the codification of meaning in terms of discursive knowledge. (Adapted from Leydesdorff (2010a), at p. 405.)



Figure 1 translates these concepts into a research design. At the lowest level (I), one can use the results of measurement. For example, one may use data from questionnaires about what a stimulus means to respondents, election polls about how one expects to vote or— closer to our lead example—relations among articles such as citations, co-authorship relations or shared co-occurrences of words (e.g., semantic maps). The distributions of observable relations contain expected information about social and cognitive structures operating upon the data. Organizing the data generates redundancy. However, without further hypotheses concerning these structures and structurations, the uncertainty in the distributions cannot be analyzed as the results of interactions among information (generating uncertainty) and meaning processing (potentially reducing uncertainty).

The structuring of the information processing is provided by the positioning of the information as in a factor model (level II). Meaning is generated in the (hypothesized) relations among information contents. Structuration is provided in terms of (next-order) codes of communication (level III). The latter can be used to relate different meanings. Whereas the positioning of the information at level II takes place at each moment of time, structuration (at level III) is based on the development of a structure over time. The operations over time can be structured because the relevant data is structured at each moment of time; these structures are reproduced and modified along trajectories.

I shall use the relations among structural components (operationalized as eigenvectors in a factor model) as a model for measuring structuration among these structural



components in the third part of this study. Does the configuration of structural components generate a synergy that feeds back as a measurable redundancy on the underlying information processing? Recent developments in entropy statistics enable us to measure this imprint of meaning processing on information processing as redundancy (Krippendorff, 2009a and b; Leydesdorff, 2009a and 2010b). But let me first turn to the specification and simulation of the mechanisms in the communication of meaning.

As noted, meaning processing (in the second contingency) cannot be measured directly. However, simulations enable us to specify the non-linear dynamics of meaning processing. Let me proceed by specifying these mechanisms of double contingency, interaction, organization, and self-organization among expectations by elaborating on algorithms from the computation of anticipatory systems (Dubois, 1998; Leydesdorff, 2008). I shall thereafter (in the third part) elaborate the empirical example of "structuration" as the generation of redundancy using intellectual (self-)organization in a set of scientific articles. The latter provide us with a textual structure. Intellectual organization can be considered as a supra-individual model of the textual domain which can leave traces within this domain.

**The theory and computation of anticipatory systems**

Reflexive systems can entertain models of (series of) events in a first contingency (*res extensa*) using a second contingency of expectations (*res cogitans*). In addition to mental models of cognitive agents, models can be exchanged among *cogitantes* in a *cogitatum*, and thus discursive knowledge can be generated and reproduced at the network level.



Note that this networked system of expectations can be reflected by human agents who perform additionally a life-cycle (in the present). The *cogitata* and the *cogitantes* can be expected to operate with different dynamics, that is, with different time perspectives.

The mathematical biologist Rosen (1985) defined *anticipatory* systems as systems which are able to entertain models. A model anticipates a future state. Dubois (1998) further distinguished between weak and strong anticipation: weakly anticipatory systems can entertain models, but strongly anticipatory systems are able to use these models reflexively for the construction of their own future states. This assumes that the models can be processed or, in other words, that meanings can be selectively exchanged and also refined. The resulting *cogitatum* thus can be expected to develop a further dynamics of its own. The model (in the *cogitatum*) is reflexively available to all the *cogitantes* who can contribute to its reconstruction reflexively.

The model provides meaning to the modeled system by anticipating a possible future state. In other words, the *cogitatum* incurs on the *cogitantes* as an expectation about possible future states. Thus, the arrow of time is reversed in the instantiation by a *cogitans*. While *cogitantes* develop recursively in historical time (the present $t$) in relation to their previous state (at $t-1$), the anticipation in the modeling subroutine (e.g., a mental model) *incurs* as a feedback on the modeled system in the present. The model advances on the modeled system by exploring possible states at a next moment of time ($t+1$), that is, hyper-incursively.



This concept of the arrow of time as a degree of freedom—unlike a given order—has been crucial to the further elaboration of the theory and computation of anticipatory systems during the last two decades. It is pertinent to the communication of meaning because meaning is provided from the perspective of hindsight, that is, incursively. First, a modeling system provides local meaning to the modeled one. Second, the communication of models adds another degree of freedom: translations among differently coded meanings become possible, and this exchange process among the expectations in a hyper-cycle can generate hyper-incursivity. Hyper-incursion presumes that the *cogitantes* are hypothesizing and communicating a next (anticipated) state of the system (at $t + 1$ or later). When the hypotheses are entertained in a discourse, discursive knowledge is generated and can provide a reconstructive feedback to the historical realization of the system in the present.

For example, *Ego*'s expectations about *Alter*'s expectations, as in the case of the above definition of "double contingency," can be considered as an example of such a hyper-incursion. *Ego* operates in the present (as $x_t$) on the basis of an expectation of its own next state ($x_{t+1}$) and the next state of an *Alter Ego* ($1 - x_{t+1}$). This can be modeled as follows:

$$x_t = ax_{t+1}(1 - x_{t+1}) \qquad (1)$$

Note that the expectation of *Alter* ($1 - x_{t+1}$) is defined in terms of *Ego*'s expectations about one's *non-Ego*; the relationship between expectations constructed in each human



mind about oneself and *Alter* precedes a possible interaction between *Ego's* and *Alter's* expectations about each other. At this level, meaning is processed in terms of an exchange between models in the mind (as *noesis*; Husserl, 1931) without implying externalization in communication (Nonaka & Takeuchi, 1995). I shall move here in a few steps towards the latter, that is, the communication of meaning (cf. Leydesdorff, 2008 and 2009b for the further elaboration of Equation 1).

Equation 1 is the hyper-incursive equivalent of the logistic equation (or Pearl-Verhulst equation) that can be used to model how a population is limited in its growth by its environment as follows:

$$x_t = ax_{t-1}(1 - x_{t-1}) \qquad (2)$$

In Equation 1, however, all references to the previous state ($t - 1$) are replaced with references to a next state ($t + 1$). Unlike the incursive variant of this same equation— $x_t = ax_{t-1}(1 - x_t)$ —for which many natural examples can be provided (Dubois, 1998; Leydesdorff, 2005), Equation 1 does not have an interpretation in the biological domain. It models a system of expectations operating upon each other, i.e., a *cogitatum*. The (self-)reference of $x_t$ to $x_{t+1}$ (in Equation 1) identifies $x_t$ as a *cogitans*.

In other words, Equation 1 captures the operation of a *cogitans* in the second contingency: the system $x$ (*Ego*) develops in the present with reference to its expectation of both its own next state ($x_{t+1}$) and the next state of its non-*Ego*, that is, ($1 - x_{t+1}$). At the



same time or concurrently, each *cogitans* incurs on its first contingency where its embodiment performs a recursive life-cycle. (This can be modeled using the incursive version of the logistic equation: $x_t = ax_{t-1}(1 - x_t)$ .[2]) The *cogitantes* can be considered as "sandwiched" between the next-order *cogitatum* and their historical manifestations in the biological domain.

Let us focus on possible interactions among the non-self-referential parts of the model in Equation 1: the term $(1 - x_{t+1})$ models *Ego*'s expectations of *Alter* as the non-*Ego*. Each *Alter Cogitans* ($y$) can be expected to entertain a similar model using $(1 - y_{t+1})$, etc. For the next-order system of interactions between these expectations of each other's expectations, these terms are the relevant parts of a model of the *communication* of meaning. In other words, the interaction can be modeled based on mutual selections of *Ego's* and *Alter's* expectations of each other. This leads to the following equation:

$$x_t = b \ (1 - x_{t+1})(1 - x_{t+1}) \qquad (3)$$

The system under study in Equation 3 can no longer be considered as a *cogitans* because it fails to include a self-referential part; only expectations are operating selectively upon each other. The new system is the result of interactions between two *cogitantes* as subroutines of a next-order interaction system which we denote again in the abstract as *x*. However, this interaction system is a social system, and no longer a psychological one.

---

[2] The incursive equation models a system which operates in the present with reference to its present environment $(1 - x_t)$ , but with also a reference to its previous state $x_{t-1}$. This equation is further elaborated (for the psychological system) in Leydesdorff (2008) and Leydesdorff & Sanders (2009).



Equation 3 can be elaborated as follows:

$$x_t = b \ (1 - x_{t+1})(1 - x_{t+1}) \tag{3}$$

$$x_t / b = 1 - 2x_{t+1} + x_{t+1}^2 \tag{3a}$$

$$x_{t+1}^2 - 2x_{t+1} + (1 - x_t / b) = 0 \tag{3b}$$

$$x_{t+1} = 1 \pm \sqrt{x_t / b} \tag{3c}$$

This interaction system can be simulated as the following oscillation:

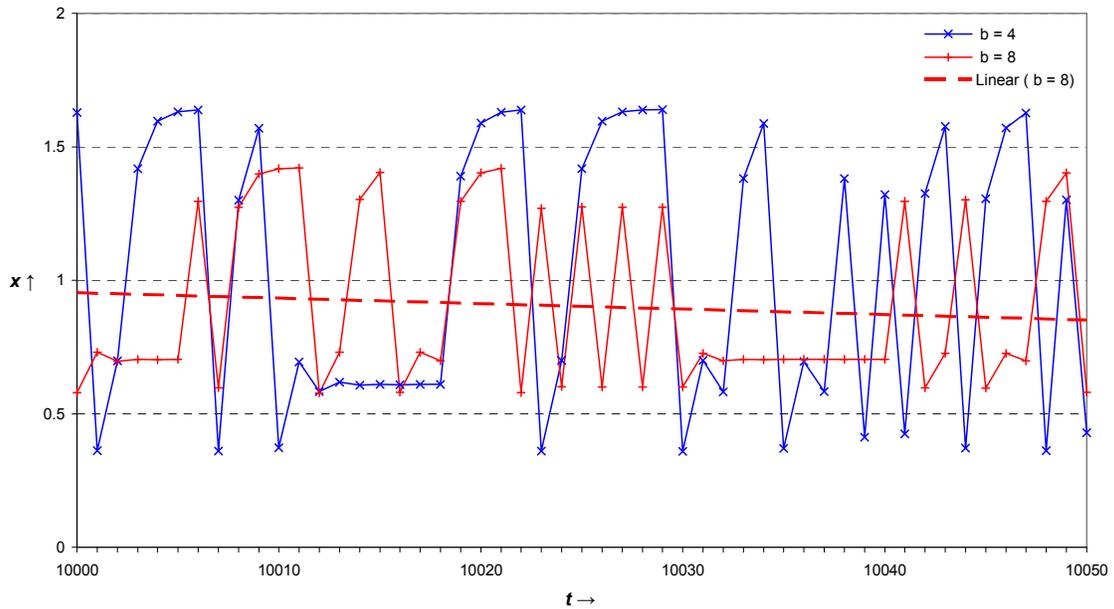

**Figure 2**: A simulation of hyper-incursive interactions.

This interaction system oscillates to variable degrees around the value of one. On average, the interaction drifts around $x = 1$ without ever reaching this value. The system



reaches its largest fluctuations (between zero and two) for $b = 2$.[3] On each side, the

interaction can be continued for a number of iterations before the alternate oscillation

resumes its operation. I modeled this here (in Excel) by using a random number to choose

the plus or minus sign in the evaluation of Equation 3c. Randomness in the variation

warrants the continuation of the interaction. In other words, interactions serve to generate

variation in the *cogitatum*. Let us now turn to organization and self-organization as next-

order operations in the communication of meaning.

**The organization and self-organization of interactions**

In this next step, Equation 3 can be extended to more complex configurations of

interactions by adding a third selection mechanism. One can add either a hyper-incursive

or incursive subroutine, and thus obtain two equations:

$$x_t = c \ (1 - x_{t+1})(1 - x_{t+1})(1 - x_{t+1}) \qquad (4)$$

$$x_t = d \ (1 - x_{t+1})(1 - x_{t+1})(1 - x_t) \qquad (5)$$

Equation 4 is a cubic equation which models a "triple contingency" of expectations

(Strydom, 1999). The equation has one real and two complex solutions.[4] Since the

---

[3] The system vanishes for $b < 2$ because the term under the root can then become larger than one, and therefore $x_{t+1} < 0$ in case of the (possibly) random choice of the minus sign in Equation 3c.

[4] The real root is of Equation 4 can be derived as (Mike Burke, *personal communication*, 10 October 2008):

$$x_{t+1} = 1 - \sqrt[3]{\frac{x_t}{c}} \qquad (4a)$$

The two complex roots are:



system cannot continue its operations further with the complex solutions, this system would if left undisturbed by other systems evolve increasingly into a single value for each value of the parameter *c*.

The parameter *c* (in Eq. 4) can be considered as a representation of the code of the communication in this "self-organizing" system. The code dampens the noise in the communication by structurating the system using a third contingency. Three contingencies operating selectively upon one another can shape a complex configuration. Note that if only a single fixed code-value *c* would operate, the routine would tend to self-organize "closure" in terms of that code. In a functionally differentiated system of communications, however, a number of values for the codes can be expected to disturb each other's tendency to operational closure. Interfaces can be expected to operate in the historical organization of communication.

Equation 5 models organization with reference to the present as an additionally incursive operation or instantiation. This equation differs from Equation 4 in terms of the time subscript of the third factor. The reference to the present in this third factor bends the system back to its present state and thus makes it historical, whereas the self-organizing system of Equation 4 and the interaction system of Equation 3 operate hyper-incursively in terms of interactions among expectations about possible future states. In Equation 5, however, the interaction among expectations is instantiated by a specific historical organization at *t* = t.

$$x_{t+1} = 1 - \sqrt[3]{\frac{x_L}{c}\left(\frac{-1 \pm i\sqrt{3}}{2}\right)}$$ (4b)



The roots of Equation 5 can be derived (analogously to Eq. 3) as follows:

$$x_t = d \ (1-x_{t+1}) \ (1-x_{t+1})(1-x_t) \qquad (5)$$

$$x_{t+1}^2 - 2x_{t+1} + 1 - x_t / [d(1-x_t)] = 0 \qquad (5a)$$

$$x_{t+1} = 1 \pm \sqrt{x_t/d(1-x_t)} \qquad (5b)$$

Simulation of this system shows that the organization of communications always vanishes after a variable number of steps for all values of the parameter $d$ (Figure 3).

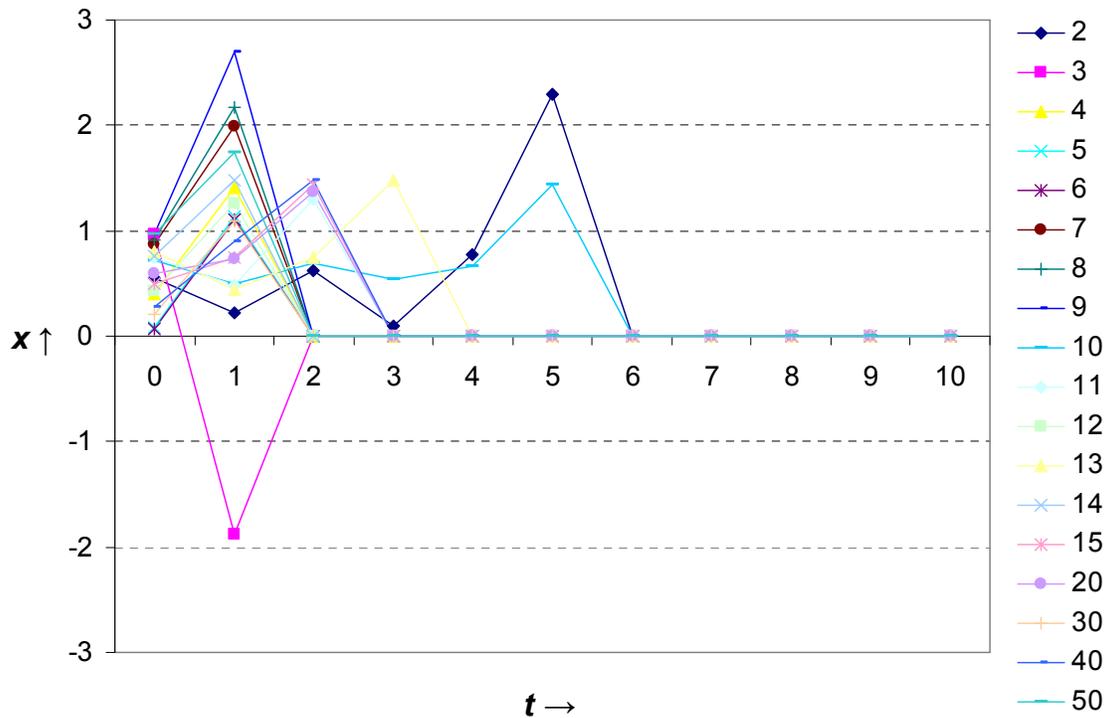

**Figure 3:** Organization of interactions for different values of the parameter $d$.



Figure 3 exhibits this development using Excel for the simulation. Excel depicts the historical end of the organization of communications as zeros, but these zeros are based on values of $x > 1$ which lead to a negative value of the denominator of the term under the root in Equation 5b. In this case, the root of this equation becomes complex and can no longer be evaluated. In other words, the organization does not disappear because of "dying," but the historical development of a specific organization can be expected to become insufficiently complex to instantiate self-organization among the fluxes of communication.

In summary, organization of communications of meaning is historical; specific organizational forms can be replaced by other organizations because of the ongoing interactions—introducing variation from below—and the hyper-incursive self-organization of the communication into codes at a relatively global level (Equation 4). Luhmann (1995, at p. 600n. [1984. at p. 551n.]) expressed this relationship among the three mechanisms in the social coordination of expectations as follows:

> "[…] in all social relations, under all circumstances a difference between society and interaction is unavoidable, but not all societies are acquainted with organized social systems. We therefore exclude organizations, but only from treatment on the level of a general theory of social systems. On the next level, that is, of concretizing the theory, one would perhaps need to distinguish between societal systems, organizational systems, and interaction systems and develop separate theories for each type because these three separate ways of forming systems (i.e., dealing with doubling contingency) cannot be reduced to one another."



Three analytically different equations (Equations 3 to 5) were derived to model these three (sub)dynamics on the basis of the initial equation (Eq. 1) which modeled double contingency as the fundamental operation. Unlike the hyper-incursive dynamics of interaction and self-organization which operate against the axis of time, organization structures communication at specific moments of time by using incursion. These instantiations also provide room for supra-individual (e.g., institutional) agency. Like (but different from) double contingency as the fundamental operation at the level of the *cogitans,* organizations can synchronously entertain different expectations because they are both both interfacing different expectations (in the first two terms of Eq. 5) and loop into the present state $x_t$ (in the third term).

Perhaps, the *cogitans* could be considered from this perspective as a minimal form of organization among expectations. Unlike the *cogitans*, however, organizations do not necessarily run an incursive routine which roots the system ($x_t$) into an historical identity with reference to a previous state $x_{t-1}$ (Equation 2). Organization of the communication of meaning can be expected to develop along a trajectory for a number of time steps, but without further input from below (variation by interaction) or regulation from above (codification of the communication), any specific organization of communications can be expected to erode in due time. The organization of communication, thus, provides us with a basis for measurement because the communication of meaning is historically instantiated. How is the possible reduction of uncertainty retained?



**The measurement of the imprint of meaning-processing**

The concept of the historical interfacing of differently coded expectations in both organizations and reflexive agents brings us back to the possibility of measurement. Historical events can be analyzed in the first contingency where (Shannon-type) information or uncertainty is processed and information theory can therefore be applied. While historical operations (along the arrow of time) necessarily generate uncertainty—the second law is equally valid for probabilistic entropy[5]—evolutionary feedback against the arrow of time may reduce the uncertainty. Providing meaning to information adds to the domain of historical events how these events could have been different. Thus, the number of possible states of the system under study is enlarged and redundancy is generated.

While at the level of each individual only thoughts and perceptions can be entertained reflexively, a model circulating in the intersubjective domain may reconstruct social reality hyper-incursively by extending the communication system(s) with new options in the model. How would a historical system entertaining a model of itself be affected by this extension of its range of possibilities? One would expect the additionally possible states (in the model) to add to the redundancy. One cannot directly measure this redundancy $R$ generated by hyper-incursive modeling because this is not taking place in the *res extensa*. However, the effect on the organization can be measured historically with the signed information measure $\mu^*$ (Krippendorff, 2009a and b; Yeung, 2009, pp. 51 ff.).

---

[5] The second law of thermodynamics holds equally for probabilistic entropy, since $S = k_B H$ and $k_B$ is a constant (the Boltzmann constant). The development of $S$ over time is a function of the development of $H$, and *vice versa*.



This information measure—unlike a Shannon measure—is signed, that is, it can be positive, negative, or zero. If $\mu^* < 0$, then uncertainty in the system is reduced because of the model entertained in the system. In other words, $\mu^*$ measures the incursion in the instantiation (Leydesdorff, 2010b). The different subdynamics (forward and backward) can be expected to operate concurrently, but they can be distinguished analytically (Figure 4).

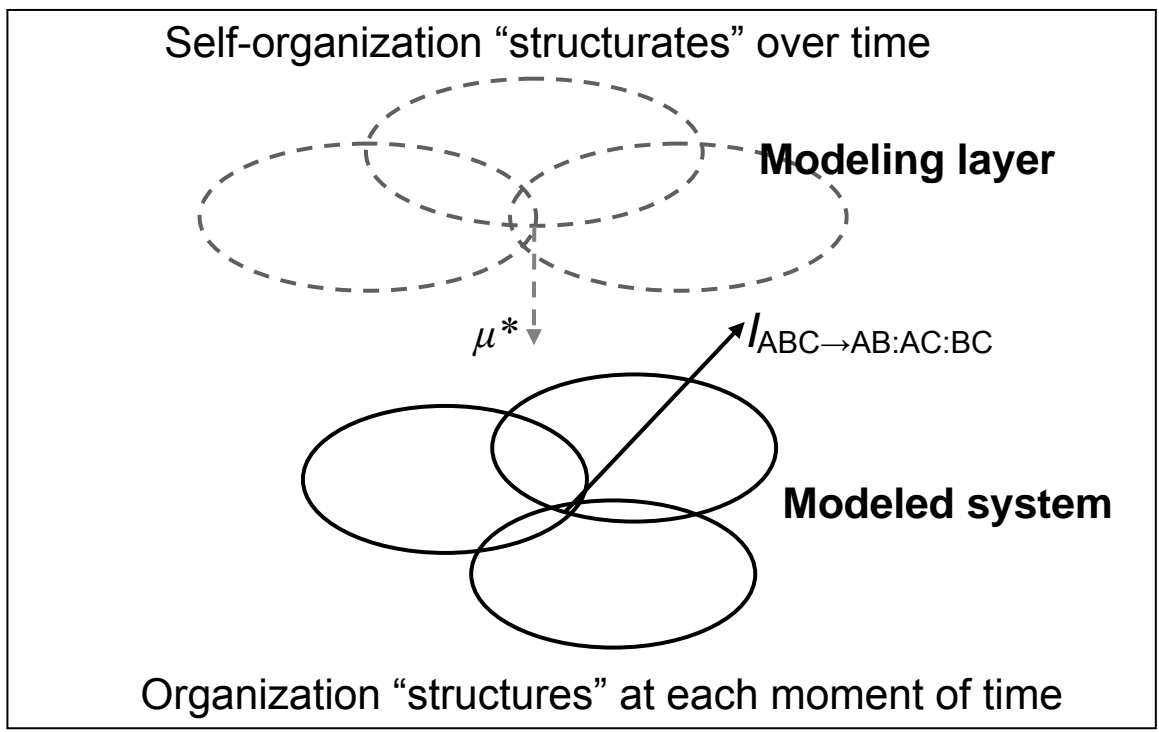

**Figure 4**: The generation of Shannon-type information ($I_{\mathrm{ABC}\rightarrow\mathrm{AB:AC:BC}}$) with time and the potentially negative feedback of mutual information ($\mu^*$) among three subsystems against the arrow of time.

This potential reduction of uncertainty can be measured as mutual information among three (or more) sources of variance. Whereas mutual information between two variables is always positive (or zero in the case of independence), mutual information among three



dimensions can also be negative and then reduce uncertainty in a system (McGill, 1954; Abramson, 1963). Yeung (2008, at pp. 51 ff.) formalized this signed measure recently as $\mu^{*}$:[6]

$$\mu^{*} = H_x + H_y + H_z - H_{xy} - H_{xz} - H_{yz} + H_{xyz} \tag{1}$$

Krippendorff (2009b) showed that this information measure should not be considered as probabilistic entropy (cf. Watanabe, 1960). In Shannon's (1948) theory the reception of a message cannot feed back on the sending of the message, and the transmission can therefore not be negative. In a follow-up study, Krippendorff (2009a) proposed taking the value of this measure as an indicator of the difference between the redundancy $R$ generated by an observer who entertains a model of the system (but is not yet informed about the historical interactions in the system), and the (Shannon-type) interaction information $I$ which he denoted as $I_{\text{ABC}\rightarrow\text{AB:AC:BC}}$ (Figure 4).

In other words, the value of $\mu^{*}$ indicates the *difference* between the redundancy ($R$) created by the modeling system and information generated by the three-way interaction information ($I$) in the modeled system. The modeled system can also be considered as an instantiation of self-organization in the sense of Equations 5 and 4, respectively. If the value of $\mu^{*}$ is measured as negative entropy, the uncertainty in the modeled system is reduced because of the modeling. If $\mu^{*} > 0$, an increase of uncertainty is indicated or—in other words—historicity prevails in the net result. As noted, the modeling within the system is based on exchanges of meaning among the meaningful communications. Thus, a synergy among the differently coded communications can be generated, but predictably to a variable extent. In other words, it is an empirical question whether observable reduction of uncertainty because of the modeling within the communication system prevails.

---

[6] Each of the terms in this formula represents a (Shannon) entropy: $H_x = -\Sigma_x\, p_x \log_2 p_x$, $H_{xy} = -\Sigma_x\, \Sigma_y\, p_{xy} \log_2 p_{xy}$, etc., where $\Sigma_x\, p_x$ represents the probability distribution of attribute $x$ and $\Sigma_x\, \Sigma_y\, p_{xy}$ the probability distribution of attributes $x$ and $y$ combined. The mutual information in two dimensions or transmission is defined as $T_{xy} = H_x + H_y - H_{xy}$.

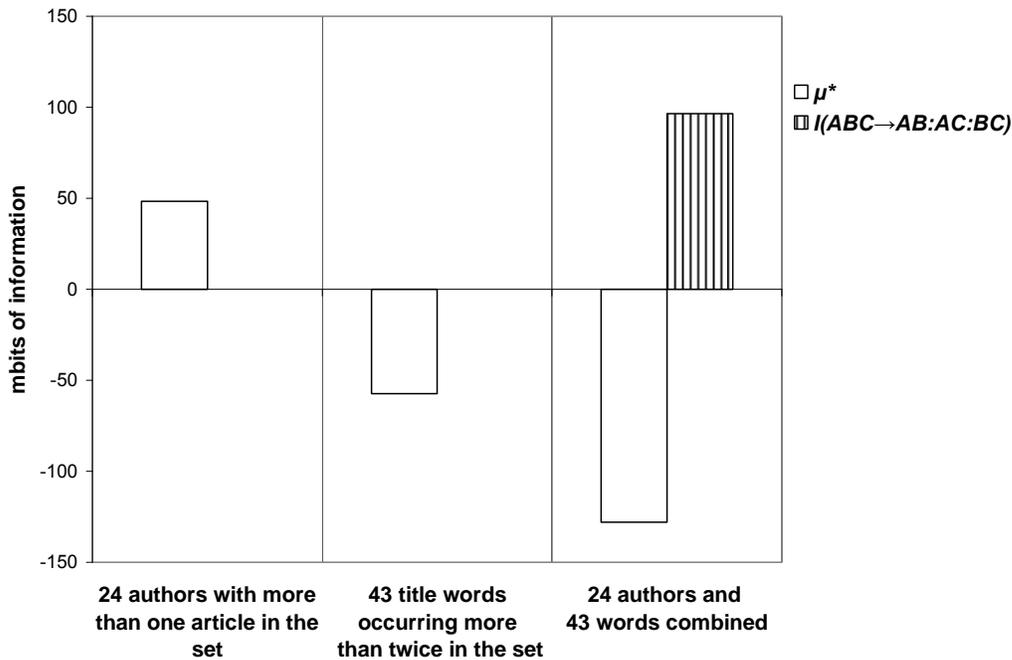

**Figure 5**: Configurational information ($\mu*$) and (Shannon-type) ternary interaction information ($I_{ABC \rightarrow AB:AC:BC}$) in the case of 24 author names and 43 title words in 102 publications in *Social Networks* 2006-2008. (Source: De Nooy & Leydesdorff, 2009).

Figure 5 shows the results of an empirical example. As can be expected, I use communications (and not agents) as units of analysis: 102 publications in *Social Networks* during the period 2006-2008 were analyzed in terms of the 43 title words



occurring more than twice and 24 author names occurring more than once in this set. The word-document and author-document matrices, respectively, were factor analyzed in order to obtain the three main components in this data. The rotated component matrices with three dimensions were subsequently input to dedicated routines to compute the values of $I_{\text{ABC}\rightarrow\text{AB:AC:BC}}$ and $\mu^*$ (De Nooy & Leydesdorff, 2009; Leydesdorff, 2010b).

In terms of the model in Figure 1, factor analysis provides us first with the structural components in the respective data matrices. This corresponds with the step between level I of *data* and level II of *structure* in Figure 1. The entropy statistics among the factor loadings provide us with a means to measure *structuration* at the next-order level III. While structure is historical and contained in the textual organization, structuration is based on next-order intellectual organization. Intellectual organization can be expected to operate in this set—representing the core of a community publishing in a specialist journal—by potentially reducing uncertainty. Intellectual self-organization at the specialty level leaves its imprint in the textual organization, and this can be measured in terms of exchanges among these texts and communalities in the sharing of textual symbols (e.g., words, author names, references, etc.; cf. De Nooy & Leydesdorff, 2009; Lucio-Arias & Leydesdorff, 2009).

Figure 5 shows that author names do *not* indicate intellectual organization of the set in terms of a negative value of the mutual information among the three main components ($\mu^* = + 48.4$ mbits). Using the matrix of title words, however, intellectual organization of the set is indicated by a negative value of $\mu^*$ (= − 57.3 mbits). Combination of these two



sets of variables shows a stronger synergetic effect among the three main components ($\mu^*$ = – 127.9 mbits) and also reveals that in this case the latent dimensions are codified to such an extent that a historical interaction term among them ($I_{ABC \rightarrow AB:AC:BC}$ = + 96.5 mbits) can also be measured (while the title words were not sufficiently codified for this effect).[7] In other words, these combined dimensions leave an imprint in the historical record. The intellectual organization, which remains volatile, is retained in the archive in terms of the combination of words and authors as markers.

These results accord with the sociological expectation: author names indicate social networks and are by themselves not sufficiently informative about intellectual organization; title words are used flexibly, but are organized intellectually (Leydesdorff, 1989 and 1997). In combination with author names, title words can be used to indicate intellectual organization at a next-order level (Callon *et al*., 1986; Leydesdorff, 2010c). Intellectual structuration ("self-organization") remains unmeasurable (since only an informed expectation), but its effects can be measured as the imprints that the structuring code leaves on the structures in the textual organization of a journal.

**Conclusions and discussion**

The two elaborations provided here to the theory of the structuration of expectations are epistemologically different. On the one side, the mechanisms operating in the *cogitata* can only be modeled and simulated, but not measured. On the other, the measurement is by definition limited to the *difference* that meaning processing in the second contingency may make for information processing in the first one, since only information processing can be measured. However, one can distinguish between the information processing itself and the feedback which the processing of information experiences from the meaning processing at a next-order level. This feedback may lead to a measurable reduction of uncertainty.

---

[7] This design can further be extended by using cited references as a third set of variables (De Nooy & Leydesdorff, 2009).

In the simulations, the meaning processing was first analytically decomposed in terms of different subdynamics such as interactions of expectations, their organization, and potential self-organization. These different operations can be expected to operate concurrently and interact in historical manifestations. On the one side, the historical organization of communications in instantiations is continuously disturbed by new interactions. These provide the variation. On the other side, organizations of communication can also be stabilized against disturbances insofar as a code of communication is operating. Self-organizing codes operating upon one another can globalize the communication in comparison to its relative stabilization in the layer of historical organizations. Globalization in the domain of meaning processing was specified as symbolic generalizations which enable us to entertain and invoke (infra-reflexively) horizons of meaning.

The simulation results are not empirical, but they enable us to clarify theoretical notions that Giddens, Luhmann, and others have developed over the last decades. Beginning with Parsons' (1951) fundamental notion of double contingency in inter-human communication, concepts like "double hermeneutics" (Giddens, 1976), "emic" *versus*



"epic" (Geertz, 1973), organization *versus* self-organization of communication (Luhmann, 1975b), and "lifeworld" *versus* "system" (Habermas, 1981) have made clear that the second *Positivismusstreit* (Adorno *et al.*, 1969) has not yet been resolved: the social sciences study not only "facts," but also, and perhaps more importantly, what these facts mean (Mulkay *et al.*, 1983; Gilbert & Mulkay, 1984).

I have recombined notably Giddens' and Luhmann's contributions because both these authors introduced abstract concepts for the operation of meaning in social systems. Giddens (1979) proposed "structuration" without specifying it otherwise than in terms of its consequences for reflexive agents and their institutions. He suggested that structuration on the basis of aggregates of actions restructures structures as "sets of rules and resources." I submit that structuration can be considered as an effect of the non-linear dynamics of interactions among agents. Interactions among interactions can lead to evolution, self-organization, and the temporary organization of expectations. Actions are historical and hence add up to trajectories, whereas expectations can also be organized into regimes of expectations or, in other words, horizons of meaning.

Luhmann proposed using the model of *autopoiesis* or self-organization, which he borrowed from neurophysiology (Maturana & Varela, 1980) but wished to apply to the self-organization of meaning in communication. However, the biological metaphor presumed a supposedly "closed" systems theory in which the functions of language (Habermas, 1987; Künzler, 1987) and reflexive agency (Giddens, 1984) cannot sufficiently be appreciated. During the 1990s, this theory increasingly became a (meta-



biological) philosophy more than a sociology (Leydesdorff, forthcoming). Self-organization in the communication of meaning can be considered as one among various subdynamics which disturb one another.

A third scholar relevant to this discussion is Pierre Bourdieu. Bourdieu (2004, at p. 78) speaks of a "transcendental unconscious" in which the knowing subject unknowingly invests by restructuring it. This "habitus" then can be considered as a historical transcendental of the subject "which can be said to be *a priori* inasmuch as it is a structuring structure which organizes the perception and appreciation of all experience, and *a posteriori* inasmuch as it is a structured structure produced by a whole series of common or individual learning processes." Although the "habitus" can be social as a role, Bourdieu's model tends to share with Giddens—and Parsons—the idea that actions and not interactions among expectations are the operators that introduce variation into the communication of meaning.

In my opinion, one advantage of Giddens' concept of "structuration" is its reference to a systems operation without identifying a "self" as in Luhmann's concept of "self-organization." Unlike Giddens, however, I did not focus on agency but on expectations, since the latter can be communicated. Thus, one can reformulate Giddens' structuration theory of action into a structuration theory of expectations. In addition to the behavioral component, action can be considered as a means to communicate expectations; expectations, however, are communicated in a *cogitatum* and are not directly observable. The communication of expectations can develop a dynamic of its own, for example,



when codes of communication emerge and thereafter can be stabilized and/or symbolically generalized. Entertaining these hypotheses enables us to study the domain of inter-human interactions without reification.

A stabilized code of communication can be expected to lead to a hierarchy in the organizational structures given a code of the communication. However, symbolically generalized codes of communication can also be differentiated, and thus provide one more degree of freedom to the communication of meaning than in the case of an identifiable system. Luhmann's concept of "self-organization" for this configuration, however, remains dependent on the biological metaphor of an identifiable "self" and functionalism. As the *cogitatum* is differentiated into *cogitata,* an identifiable self of the social system can no longer be expected. In other words, the *cogitatum* should not even metaphorically be considered as a *cogitans.* Therefore, I prefer Giddens' concept of "structuration" which adds the structured dynamics of expectations to their structuring at each moment of time. Unlike Giddens, however, this is not a structuration theory of action, but a structuration theory of expectations.

What has thus been gained since Husserl's (1935/36) *Crisis*? In my opinion, the epistemological order of priority changed with Husserl's reflection: the intersubjective *cogitatum* provides a necessary—Husserl would say "transcendental"—condition for the subjective *cogito* although the latter remains a historical condition for the former.[8]

---

[8] Künzler (1987, at p. 331) formulated that "between Luhmann's marginalization of language (cf. Habermas, 1985, at p. 438) and Habermas's foundation of sociology in the theory of language, one should be able to find the comparatively innocent consideration of meaning as the *ratio essendi* of language and language as the *ratio cognoscendi* of meaning" (my translation). Our argument, however, goes beyond this



Interhuman communication transcends individual consciousness. When the codes in the communication can also be generalized symbolically, they begin to structure horizons of meaning. Husserl used the word 'rooting' for the historical origins of the *cogitatum* in human reflexivity, but he emphasized that an intentionality that transcends the individual can be developed further.

Husserl (1929, at p. 138) regretted that he could not specify the mental predicates for studying this "layer of meaning which provides the human world and culture, as such with a specific meaning" other than as an analogy to the categories of philosophical reflection by the *cogito*, that is as a (meta-psychological) transcendency. I submit that further elaborations of information theory and the theory and computation of anticipatory systems have provided us with categories for studying the evolution of the *cogitata* as inter-human coordination systems.

### Acknowledgement


I am grateful to Sally Wyatt for discussions about previous drafts.

---

position because we argue that codified knowledge in a functionally differentiated configuration can only be analyzed by invoking codes of communication symbolically generalized at a level beyond human language. For example, we need not words but money or credit for economic transactions.